\documentclass[traditabstract]{aa}

\usepackage{graphicx}
\usepackage{multirow}
\usepackage{natbib}
\usepackage{adjustbox}
\usepackage{txfonts}
\usepackage{mathrsfs}
\usepackage[colorlinks = true,linkcolor = blue,urlcolor = blue,citecolor = blue]{hyperref}
\usepackage{xcolor}
\definecolor{yellowgray}{rgb}{0.90, 0.90, 0.2}
\definecolor{bluegray}{rgb}{0.20, 0.60, 0.80}
\definecolor{palered}{rgb}{0.99, 0.40, 0.5}
\definecolor{darkgray}{rgb}{0.35, 0.35, 0.35}
\definecolor{darkgrayb}{rgb}{0.75, 0.75, 0.75}
\definecolor{palegray}{rgb}{0.96, 0.96, 0.96}

\begin{document} 

\title{DeSIRe: Departure coefficient aided Stokes Inversion based on
  Response functions}
  \titlerunning{DeSIRe inversion code}
   \author{B. Ruiz Cobo \inst{1,2} \and C. Quintero Noda \inst{1,2,3,4,5}\thanks{Corresponding author}  \and R. Gafeira\inst{6} \and H. Uitenbroek\inst{7}  \and D. Orozco Su\'arez\inst{8}  \and  E. P\'aez Mañ\'a\inst{1,2}
             }
   \institute{ Instituto de Astrof\'isica de Canarias, E-38200, La Laguna, Tenerife, Spain.\\
              \email{carlos.quintero@iac.es}     
\and     Departamento de Astrof\'isica, Univ. de La Laguna, La Laguna, Tenerife, E-38205, Spain
\and
Rosseland Centre for Solar Physics, University of Oslo, P.O. Box 1029 Blindern, N-0315 Oslo, Norway
\and
Institute of Theoretical Astrophysics, University of Oslo, P.O. Box 1029 Blindern, N-0315 Oslo, Norway
\and
Institute of Space and Astronautical Science, Japan Aerospace Exploration Agency, Sagamihara, Kanagawa 252-5210, Japan
\and   Univ Coimbra, IA, CITEUC, OGAUC, Coimbra, Portugal
\and    National Solar Observatory, University of Colorado Boulder, 3665 Discovery Drive, Boulder, CO 80303, USA
\and    Instituto de Astrof\'isica de Andaluc\'ia (CSIC), Apdo. de Correos 3004, E-18080 Granada, Spain\\
             }
   \date{Received 25 March 2021 ; accepted 28 January 2022  }

% \abstract{}{}{}{}{} 
% 5 {} token are mandatory
 
\abstract{Future ground-based telescopes, such as the 4-metre class facilities
  DKIST and EST, will dramatically improve on current capabilities for
  simultaneous multi-line polarimetric observations in a wide range of
  wavelength bands, from the near-ultraviolet to the near-infrared. As
  a result, there will be an increasing demand for fast diagnostic
  tools, i.e., inversion codes, that can infer the physical properties
  of the solar atmosphere from the vast amount of data these observatories
  will produce. The advent of substantially larger apertures,
  with the concomitant increase in polarimetric sensitivity, will drive
an increased interest in observing chromospheric spectral lines.
  Accordingly, pertinent inversion codes will need to take
  account of line formation under general non-local thermodynamic equilibrium (NLTE) conditions. Several currently available codes can
  already accomplish this, but they have a common
  practical limitation that impairs the speed at which they can invert
  polarised spectra, namely that they employ numerical evaluation of
  the so-called response functions to changes in the atmospheric
  parameters, which makes them less suitable for the analysis of very
  large data volumes. Here we present DeSIRe (Departure coefficient aided Stokes Inversion
  based on Response functions), an inversion code that integrates the
  well-known inversion code SIR with the NLTE radiative transfer
  solver RH. The DeSIRe runtime benefits from employing analytical
  response functions computed in local thermodynamic equilibrium (through SIR), modified with
  fixed departure coefficients to incorporate NLTE effects in
  chromospheric spectral lines. This publication describes the
  operating fundamentals of DeSIRe and describes its behaviour,
  robustness, stability, and speed. The code is ready to be used by the
  solar community and is being made publicly available.
}


\maketitle

\section{Introduction}

Our knowledge of the physics of the Sun comes from the analysis of the radiation we receive from it. Atoms and molecules that appear with different concentrations in the solar atmosphere interact with that radiation, leaving a characteristic imprint in the form of so-called spectral lines. Those spectral lines are sensitive to perturbations in the plasma medium they belong to, showing specific signatures due to differences in, for example,  the temperature, velocity, or magnetic field. Thus, we can extract information about the solar atmosphere and all the phenomena that occur in it through the analysis of spectropolarimetric observations of various spectral lines.

Interestingly, inferring the depth dependence of physical properties of the solar
atmosphere from spectropolarimetric observations (i.e. inverting the
spectrum) typically requires solving the radiative transfer equation
(RTE) many times and minimising the difference between
observed and modelled spectra. Thus, there is a strong incentive to
define inversion mechanisms that both minimise the number of calls to
the RTE solver and make this solver as efficient as
possible. The degree of complexity in inversion codes depends on the
amount of physics implemented to represent the atmosphere underlying
the observations as closely as possible. Inversion codes can
  go from simplified models, such as the widely used Milne-Eddington
  approximation \citep[for instance, ][]{Skumanich1987}, to more
  complex ones that account for the full radiative transfer problem
  under local thermodynamic equilibrium (LTE) conditions
  \citep[see][for a
    review]{delToroIniesta2016,delaCruzRodriguez2017}. In this
context, the Stokes Inversion based on Response functions (SIR) code
\citep{RuizCobo1992} is, among others, of particular
interest. It solves the RTE in LTE and employs response functions
\citep[RFs; e.g. ][]{Landi1977}, which are evaluated analytically. This approach makes it fast in comparison with other codes. However,
LTE is mainly suitable for photospheric line modelling and is not
applicable for lines that form in the chromosphere, where general
non-local thermodynamic equilibrium (NLTE) conditions prevail.

Several inversion codes that can deal with the NLTE problem are
already available. In the particular case of optically thick media,
the Non-LTE Inversion COde using the Lorien Engine (NICOLE; \citealt{SocasNavarro2000,SocasNavarro2015}) was the first such
full-Stokes inversion code available for the solar community. NICOLE
works in NLTE and complete redistribution (CRD) and can invert the
Stokes profiles of any given transition. Unfortunately, it only deals
with single atomic species, so its applicability to multi-line
observations involving different species is limited. Moreover, the
code evaluates the RFs numerically to incorporate NLTE
effects, although it is based on SIR. More recently, the STockholm
inversion Code \citep[STiC;][]{delaCruzRodriguez2016,delaCruzRodriguez2019} was published. It is
based on the NLTE forward solver RH \citep{Uitenbroek2001,Uitenbroek2003} and adds several improvements to
that code. For example, STiC solves the RTE using cubic Bezier
solvers \citep{delaCruzRodriguez2013}, allows the inversion of lines
from multiple atomic species, and includes partial re-distribution (PRD)
effects. Similarly to the case of NICOLE, it uses numerical RFs in its
minimisation scheme. The main disadvantage when using numerical RFs is
the code runtime. Since within the inversion process it is necessary
to evaluate the response of the Stokes profiles to changes in
all physical quantities in the
model (i.e. compute the derivatives of the Stokes profiles with respect to
the atmospheric parameters) at the designated nodes,
the total time per inversion iteration
increases considerably. However, both NICOLE and STiC take advantage of
modern compute clusters to speed up the analysis of large
observational datasets by employing parallelisation.

In a similar time frame as when STiC was developed,
\cite{Milic2017,Milic2018} presented an alternative NLTE inversion
code, SNAPI (from Spectropolarimetic NLTE Analytically Powered
Inversion), which is based on the analytical\footnote{It is important
to bear in mind that none of the codes presented here infer the RFs
fully analytically from the mathematical point of view since the
solution of the RTE is always a numerical problem. Only in simplified
atmospheres, as in a Milne-Eddington approximation, this would be
strictly valid. Hence, we refer to numerical RFs when they are
computed using finite difference methods.} approximation of the NLTE
RFs. The main improvement of SNAPI over STiC and NICOLE is the reduced
time for computing the RFs, which are calculated along with the formal
solution of the RTE problem. Similarly to NICOLE, however, SNAPI only
inverts for one atomic species at a time and employs the CRD approximation.

To improve on these current inversion codes, we present in this
paper a novel implementation to speed up the NLTE
inversion problem, combining the well-established SIR and RH
codes. The underlying concept is that, instead of solving the full
inverse NLTE problem, which requires solving the statistical
equilibrium equations and calculating the RFs with full NLTE
sensitivities repeatedly, we approximate the level populations in the
required RFs by introducing the so-called fixed
departure coefficient (FDC) approximation, similarly to what was
initially proposed for NICOLE in \cite{SocasNavarro2000}. Departure
coefficients are the ratio of the population of a given atomic level
under NLTE conditions over that in LTE. Moreover, we use exact analytical
LTE RFs modified by the FDCs to approximate the response of NLTE lines
to changes in the atmospheric physical quantities. This code, which we have named
the Departure coefficient aided Stokes Inversions based on Response
functions (DeSIRe), expands the inversion
engine of SIR by taking the NLTE forward solution and the NLTE
departure coefficients from the RH code. The main assumption is that
LTE analytic RFs modified with FDCs are
accurate enough for computing the sensitivity of the NLTE Stokes
profiles to model perturbations. The gain of this new approach is
twofold: the inversion time decreases and the analysis of both LTE and NLTE lines is simplified. The
upcoming large amount of observational data from ground-based
telescopes such as the Daniel K. Inoue Solar Telescope \citep[DKIST;][]{Rimmele2020} and the European Solar Telescope
\citep[EST;][]{Collados2013}, estimated to be of the order of petabytes per
year, should undoubtedly benefit from the reduction in computation time provided by DeSIRe.

In Sect.\ \ref{sec:code} we describe how we use the FDCs to construct an inversion code that is suitable for
analysing polarimetric spectra of chromospheric lines formed under
general NLTE conditions as well as LTE lines. We present several test
results of inversions of forward modelled spectra from a snapshot of a
realistic radiation magneto-hydrodynamic (Rad-MHD) simulation, which
includes chromospheric physics, in Sect.\ \ref{inver3D_sec}. Finally,
we conclude in Sect.\ \ref{sec:discussion} that the code is stable,
reliable, accurate, and fast when employed in inversion experiments based on the realistic simulations of the solar magneto-convection employed
here. Therefore, it will be a valuable tool for interpreting the large
amount of data that will be produced by upcoming 4-metre class
ground-based facilities such as DKIST and EST, as well as by future flights of
the Sunrise balloon.

\section{Equations of radiative transfer and statistical equilibrium\label{sec:RTE}}

\subsection{Assumption of a stationary and plane-parallel atmosphere}

The equation that describes how radiation travels through a
medium is known as the RTE \citep[see,
for instance,][]{Mihalas1978}. In this work, we focus on the
specific case of radiation coming from a stellar atmosphere that is
geometrically sufficiently thin compared with the stellar radius; in other words,
we employ the so-called plane-parallel approximation. We also assume the
medium is stationary.

If we suppose that a spectral line of interest has a rest wavelength
$\lambda_1$ and that it is blended with several spectral lines
$\lambda_b$, with $b=2...n_b$, the bound-bound (line) absorption
happens at a narrow wavelength range: the function describing the
wavelength dependence of the absorption is named the absorption profile
and is commonly expressed normalised in area. That absorption profile
represents the probability that a photon is absorbed in a given
wavelength close to the central one. We denote
$\varphi^b(\lambda-\lambda_b)$ the absorption profile of the $b$
component, evaluated at wavelength $\lambda$. For the emission
process, $\psi^b(\lambda-\lambda_b)$ designates the emission
profile. In PRD, when the emission profile depends in part on
coherently scattered photons the emission and absorption profiles
will generally be different from each other, while in CRD they are
assumed to be the same \citep[see more in][]{Mihalas1978}. Complete redistribution is
adequate for all photospheric and most chromospheric lines. It works
well for transitions of intermediate strength, such as the \ion{Ca}{ii}
infrared triplet lines \citep[see, among others,][]{Uitenbroek2006,delaCruzRodriguez2012,QuinteroNoda2016},
but not for the strongest chromospheric lines, such as \ion{H}{i} $\alpha$
and $\beta$, the \ion{Ca}{ii} H and K lines, and the \ion{Mg}{ii}
$h$ and $k$ lines, which all require treatment with PRD.

The shape of both emission and absorption profiles results from several
mechanisms, called broadening mechanisms. Among the most important
ones are the natural, Doppler, collisional, and Stark broadening. In
general we can calculate $\varphi^b(\lambda-\lambda_b)$ (and
$\psi^b(\lambda-\lambda_b)$) as the convolution of a Gaussian and a
Lorentzian function. The former has a width ($\sqrt{2}$ times the
standard deviation) equal to de Doppler width, $\Delta \lambda_D$, that
we can define as
\begin{equation}
\Delta \lambda_D=\lambda_b/c  \sqrt{\xi^2+\frac{2kT}{M}},
\end{equation}
where $\xi$ is the microturbulent velocity, $T$ the
temperature, $M$ the mass of the atom involved in the transition, and
$c$ and $k$ have their usual meanings. The Lorentzian function is
defined by the parameter $a$, usually called damping parameter:
\begin{equation}
a=\Gamma \lambda_b^2/(4 \pi c \Delta \lambda_D),
\end{equation}
with $\Gamma$ being the damping constant, which is determined
considering the sum of collisional and natural broadening mechanisms.

The result of convolving the Gaussian and the Lorentzian functions is
named the Voigt function, $H(a,v)$ \citep[][]{Landi1976}:
\begin{equation}
  H(a,v)= a/\pi \int_{-\infty}^{\infty} exp(-y^2)((v-y)^2+a^2)^{-1} dy,
  \label{eq:voigt}
\end{equation}
where $v$ is the wavelength in Doppler width units:
\begin{equation}
v= (\lambda-\lambda_b)/\Delta \lambda_D.
\end{equation}

Following the previous publication, we can define the Faraday-Voigt
function  $F(a,v)$ as well, which we will use later on:\begin{equation}
 F(a,v)= 1/(2\pi) \int_{-\infty}^{\infty} exp(-y^2)(v-y)((v-y)^2+a^2)^{-1} dy.
 \label{eq:faradayvoigt}
\end{equation}

An absorption event could happen by continuum or line processes. We
can define the absorption coefficient $k_{\lambda}$ (i.e. the fraction
of $I_{\lambda}$ absorbed per unit of length) as
\begin{equation}
  k_{\lambda}(\lambda)  = k_c (\lambda)+ \sum_{b=1}^{n_b} k^b_l
  \varphi^b(\lambda-\lambda_b), 
   \label{abs1}
\end{equation}
with $k_c$ and $k^b_l$ being the contribution of continuum,
and line absorption coefficients of the $b$ component, respectively.

Similarly, the emission has a contribution of both continuum and line. Thus,
we can define the emission coefficient $\eta_{\lambda}$ as
\begin{equation}
  \eta_{\lambda}(\lambda) \equiv k_c B_{\lambda}(\lambda) +
  \sum_{b=1}^{n_b} k^b_l S^b \psi^b(\lambda-\lambda_b),
   \label{eta1}
\end{equation}
where $B_{\lambda}$ is the Planck function and $S^b$ the line source
function, which is defined as the ratio of the emission over the
absorption coefficient. With all these ingredients we can write the RTE as
\begin{equation}
   \frac{dI_{\lambda}}{dz}  = - k_{\lambda} I_{\lambda} + \eta_{\lambda}.
   \label{RTE1}
\end{equation}
We can understand the previous equation (i.e. the RTE) as an
expression of the energy equation. In a plane-parallel and stationary
medium, the change of specific intensity $I_{\lambda}$ (power per
square cm, unit of wavelength and steradians) of a beam crossing a
layer $dz$ decreases by absorption processes and increases by emission
events. Usually, scattering is considered as an absorption process and
the stimulated emission as a negative absorption.

It is customary to write the thickness of the layer, $dz$, in terms of the
length of the free path of a photon at a given fixed reference
wavelength, usually that of the continuum at 500~nm.
Following this tradition we define the continuum optical
depth at 500~nm $\tau_{500}$, following Eq. \ref{RTE1}, by
\begin{equation}
   \mathrm{d} \tau_{500}= -k_{500} \mathrm{d} z,
   \label{opticaldepth}
\end{equation}
which represents the number of photon mean free paths at 500~nm
over the geometrical height interval $\mathrm{d} z$.

Dividing Eq. \ref{RTE1} by Eq. \ref{opticaldepth} and
defining the source function $S_{\lambda}$ as the emission over a mean
free path, 
\begin{equation}
S_{\lambda}=\frac{\eta_{\lambda}}{k_{\lambda}},
\end{equation}
we can get the following expression for the RTE:
\begin{equation}
  \frac{dI_{\lambda} }{d\tau_{500}}  = \frac{k_{\lambda}}{k_{500}} \left(
  I_{\lambda} -S_{\lambda}  \right).
   \label{RTE2}
\end{equation}
The rationale for this transformation is that both $k_c$ and $k_l$
(and consequently $k_{500}$) are proportional to the mass density,
which is, in many cases, unknown. Additionally, in some cases, the
source function can be described in a simplified way. For instance
under the LTE approximation, $S_{\lambda}$ becomes equal to the Plank
function $B_{\lambda}$.

However, we will see that in the case of polarised light, the absorption
coefficient $k_{\lambda}$ becomes a matrix. Consequently, to avoid
evaluating the inverse of the absorption matrix, it is customary to
define the source function per continuum optical depth interval,
\begin{equation}
\mathscr{S_{\lambda}} =\frac{\eta_{\lambda}}{ k_{500}},
\end{equation}
that is, the emission through a mean free path of a continuum photon at
$500$ nm, instead of through a mean free path of a photon at
wavelength $\lambda$. The RTE then becomes
\begin{equation}
  \frac{dI_{\lambda} }{d\tau_{500}}  = \frac{k_{\lambda}}{k_{500}}
  I_{\lambda} -\mathscr{S_{\lambda}}.
   \label{RTE3}
\end{equation}

\subsection{Radiative transfer equation for polarised light}

We can find a description of the RTE for polarised light in many
books and papers
\citep[among others, ][]{Landi1985,Landi2004,delToroIniesta2003}.
The RTE can be
written, for the particular case of a polarised light beam going
through a stationary plane-parallel atmosphere, as a generalisation
of Eq. \ref{RTE3}:
\begin{equation}
   \frac{d\vec{I}}{d\tau_{500}}  = \mbox{K} \vec{I}-\vec{\mathscr{S}},
   \label{RTEpol}
\end{equation}
where $\vec{I}=(I,Q,U,V)$ stands for the Stokes pseudo-vector,
$\bf{\mathscr{S}}$ for the Source function per continuum unit vector,
and $\mbox{K}$ corresponds to the absorption matrix, which can be
written as
\begin{equation}
  \mbox{K} \equiv \frac {k_c}{k_{500}} \mbox{1} + \sum_{b=1}^{n_b}
  \frac{k^b_l}{k_{500}}\mbox{K}^b
   \label{Keq1}
.\end{equation}
The matrix $\mbox{1}$ is the 4$\times$4 identity matrix, and the array
$\mbox{K}^b$ is given in terms of only seven elements \citep[for
  example,][]{Landi1981}:
\begin{equation}
        \mbox{K}^b= \left (\begin{array}{cccc}
    \eta_{I} & \eta_{Q} & \eta_{U} & \eta_{V}\\
    \eta_{Q} & \eta_{I} & \rho_{V} & -\rho_{U} \\
    \eta_{U} & -\rho_{V} & \eta_{I} & \rho_{Q}\\
    \eta_{V} & \rho_{U} & -\rho_{Q} & \eta_{I}\end{array} \right),
\label{Keq2}
\end{equation}
where the different elements inside the matrix are given by
\begin{eqnarray}
  \eta_{I} = \frac{1}{2}\left [\phi_{p}sin^{2}\gamma + \frac{1}{2}(\phi_{b} +
    \phi_{r})(1 + cos^{2}\gamma) \right]\\
  \eta_{Q} = \frac{1}{2}\left [\phi_{p} - \frac{1}{2}(\phi_{b} +
    \phi_{r})\right]sin^{2}\gamma cos2\chi\\
  \eta_{U} = \frac{1}{2}\left [\phi_{p} - \frac{1}{2}(\phi_{b} +
    \phi_{r})\right]sin^{2}\gamma sin2\chi\\
\eta_{V} = \frac{1}{2}(\phi_{r} - \phi_{b})cos\gamma\\
\rho_{Q} = 1\left [\Psi_{p} - \frac{1}{2}(\Psi_{b} +
  \Psi_{r})\right]sin^{2}\gamma cos2\chi\\
\rho_{U} = 1\left [\Psi_{p} - \frac{1}{2}(\Psi_{b} +
  \Psi_{r})\right]sin^{2}\gamma sin2\chi\\
\rho_{V} = \frac{1}{2}(\Psi_{r} - \Psi_{b})cos\gamma,
\label{etas}
\end{eqnarray}
$\gamma$ and $\chi$ being the inclination and azimuth angles of the
magnetic field vector, respectively. Profiles $\phi_{j}$ ($j = p, b, r$) are
the absorption profiles for polarised light. The indices $p, b$, and $r$
represent the $\pi$, blue $\sigma$, and red $\sigma$ Zeeman
components. Profiles $\Psi_{j}$ ($j = p, b, r$) are the dispersion profiles and
represent the magneto-optical effects, that is, the variations in the
polarisation state induced by light passing through dichroic media
\citep[more details in][]{Landi1985}. Profiles $\phi_{j}$ and $\Psi_{j}$
are described by the Voigt (Eq.\ \ref{eq:voigt})
and Faraday-Voigt (Eq.\ \ref{eq:faradayvoigt}) functions, respectively,
as follows:
\begin{align}
\phi_{j} &= \sum_M S_M^j H(a,v-v_M^j)\\
\Psi_{j} &= 2\sum_M S_M^j F(a,v-v_M^j).
\end{align}
\citet{Landi1976} presented expressions to evaluate the Zeeman shift
$v_M^j$ of each component $M$ in terms of the magnetic field and the
strength $S_M^j$ of that component.
The Source function per continuum unit vector can be expressed as
\begin{equation}
  \vec{\mathscr{S}} = (\eta_{I}\mathscr{S}_\lambda,\eta_{Q}\mathscr{S}_\lambda,
  \eta_{U}\mathscr{S}_\lambda, \eta_{V}\mathscr{S}_\lambda).
\end{equation}

Finally, lower level populations $n^b_{low}$ enter in the evaluation
of absorption matrix (Eq.\ \ref{Keq1}) multiplying the absorption
coefficients $k^b_l$, and also together with the upper level
population $n^b_{up}$ in the source function through terms $k^b_l$ and
$S^b$.

%HU
\subsection{Radiative transfer under NLTE conditions\label{sec:NLTE}}
Excitation and de-excitation, as well as ionisation and recombination
processes between atomic levels can take place in two ways: by
collisions (most importantly with free electrons, as they are lighter
and have, on average, much larger thermal velocities than atoms and
ions), or by absorption and emission of a photon. In deeper, denser
layers of the solar atmosphere, like the photosphere, collisions
dominate. Going upwards, collisional probabilities become less and less
important with the exponential drop in density in a gravitationally
stratified atmosphere, while radiative probabilities are independent
of density. As a result, population numbers, and with them, the
opacity and emissivity, are determined more and more by the radiation
field, with increasing height. Thus, population numbers in one
location in the atmosphere become dependent on the radiation field
originating locally as well as from many different other locations in
the atmosphere.  As a result, we have to solve for NLTE.

Assuming an atomic species $A$ with energy levels $i, j = 1,\ldots , N_{A}$,
we define the quantities $V$ and $U$ at frequency $\nu$ and
direction $\vec{n}$ for a
bound--bound transition between levels $i$ and $j$ with $j > i$ as
\begin{align}
  V_{ij} &= (h\nu / 4\pi) B_{ij} \phi_{ij}(\nu, \vec{n}),\\
  V_{ji} &= (h\nu / 4\pi) B_{ji} \psi_{ij}(\nu, \vec{n}),\\
  U_{ji} &= (h\nu / 4\pi) A_{ji} \psi_{ij}(\nu, \vec{n})
\end{align}
\citep[cf., ][]{Rybicki+Hummer1992, Uitenbroek2001}.
The quantities $B_{ij}, B_{ji}$, and $A_{ji}$ are the Einstein
coefficients for absorption, stimulated emission and spontaneous
emission, respectively,
and $\phi$ and $\psi$ are the line absorption and emission
profiles at frequency $\nu$ and in the direction
$\vec{n}$.

Similarly, for a bound--free transition $(i, j)$, the
quantities $V$ and $U$ are defined as
\begin{align}
  V_{ij} &= \alpha_{ij}(\nu),\\
  V_{ji} &= n_e \Phi_{ij}(T) e^{-h\nu/kT} \alpha_{ij}(\nu),\\
  U_{ji} &= n_e \Phi_{ij}(T)\left(\frac{2h\nu^3}{c^2} \right)
    e^{-h\nu/kT} \alpha_{ij}(\nu),
\end{align}
where $\alpha_{ij}(\nu)$ is the photoionisation cross-section and
$\Phi_{ij}(T)$ is the Saha--Boltzmann function:
\begin{equation}
  \Phi_{ij}(T) = \frac{g_i}{2g_j} \left( \frac{h^2}{2 \pi kT} \right)^{3/2}
  \exp{\left[ (E_j - E_i)/kT \right]}.
\end{equation}
Here $n_e$ and $T$ are the electron density and temperature, respectively,
and $g_i$ and $g_j$ are the statistical weights of the lower and upper levels.
Given these definitions for $V$ and $U$, we can now write the radiative rate
$R_{ij}$ of a transition from level $i$ to $j$ as an integral over
frequency and all solid angles of specific intensity $I(\nu, \vec{n})$:
\begin{equation}
  R_{ij} = \oint \mathrm{d} \Omega \int \frac{\mathrm{d} \nu}{h \nu}
  \left\{ U_{ij}(\nu, \vec{n}) +  V_{ij}I(\nu, \vec{n}) \right\},
  \label{eq:radrate}
\end{equation}
with the understanding that $U_{ij} \equiv 0$ if $i < j$.

Formally, in each location of the atmosphere, the population numbers 
$n^{A}_{i}$ of atomic species $A$
are given by the set of statistical equilibrium equations for each $i$:
\begin{align}
  \frac{\mathrm{d} n_i}{\mathrm{d} t} & = n_i
  \left( \sum_{i=1}^{N_A}  R_{ij} + C_{ij} \right) -
  \sum_{i=1}^{N_A} n_j \left(  R_{ji} + C_{ji} \right),\label{eq:stateq1}\\
  \sum_{i=1}^{N_A} n_i &= n^A_{\mathrm{tot}} = A_A n_H, \label{eq:stateq2}
\end{align}
where $R_{ij}$ and $C_{ij}$ are the radiative and collisional
transition rates between levels $i$ and $j$, respectively,
$n^A_{\mathrm{tot}}$ is the total number density of atom $A$, $A_A$ is
the abundance of atomic species $A$, and $n_H$ the number density of
hydrogen. During inversions it is assumed the atmosphere is stationary, so that
$\mathrm{d} n_i / \mathrm{d} t = 0$. Equation \ref{eq:stateq2} is the
abundance equation. It follows from Eq. \ref{eq:stateq1} that,
in general, the occupation numbers $n_i$ are determined by the balance
between the rates of excitation and de-excitation and ionisation and
recombination, through radiative as well as collisional processes.

When collisions dominate in the statistical equilibrium equation, Eq.
\ref{eq:stateq1},
detailed balance holds between each pair of levels $i,j$ \citep[cf.,
][]{Mihalas1978}, and the population numbers are given by their LTE
values $n^*$:
\begin{align}
  n_i^* C_{ij} &= n_j^* C_{ji},\\
  \sum_{i=1}^{N_A} n_i^* &= n^A_{\mathrm{tot}.}
\end{align}
As a result, we can use LTE opacities and source functions at the
local kinetic temperature, $T$, for lines and continua that have their
formation layers in the deeper, denser layers of the
atmosphere. However, in higher layers we have to solve explicitly and
simultaneously for the radiation field, via the equation of transfer,
and the statistical equilibrium equations.

The dependence of the set of statistical equilibrium equations on the
radiation field makes the solution of these equations a non-local and
non-linear problem, because the radiation field at one location depends
on the absorption and emission coefficients in other parts of the atmosphere,
and these coefficients depend in turn on the population numbers in those
locations. This globally coupled non-linear problem can only be solved
iteratively, which is what codes like RH code accomplish
\citep{Uitenbroek2001}.

%HU

\subsection{Departure coefficients}

For each transition $b$ between two levels, the departure coefficients
of the upper (or lower) level correspond to the ratio between the
actual population of this level and that population evaluated in LTE:
\begin{equation}
\beta^b_{low,up}=\frac{n^b_{low,up}}{n^{b*}_{low,up}}.
\end{equation}
Introducing departures coefficients, we can rewrite Eq. \ref{Keq1} as
\begin{equation}
  \mbox{K} \equiv \frac {k_c}{k_{cr}} \mbox{1} + \sum_{b=1}^{n_b}
  \frac{k^{*b}_l \beta^b_{low}}{k_{cr}}\mbox{K}^b,
   \label{Keq2}
\end{equation}
where $k^{*b}_l$ is the line absorption coefficient of the $b$
component evaluated in LTE.
We can rewrite Eq. \ref{eta1} as
\begin{equation}
  \eta_{\lambda}(\lambda) \equiv k_c B_{\lambda}(\lambda) +
  \sum_{b=1}^{n_b} k^{*b}_l \beta^b_{low} S^b \psi^b(\lambda-\lambda_b)
   \label{eta2}
\end{equation}
\noindent and the contribution of the line source function $S^b$ as
\begin{equation}
  S^b=\frac {2 h c^2}{\lambda_b^5}
  \frac {1}{\beta^b_{low}/\beta^b_{up} \exp{(hc/\lambda_b k T)} -1}
   \frac{\psi^b}{\varphi^b}.
    \label{Seq2}
\end{equation}

%%%%%%%%%%%%%%%%%%%%%%%%%%%%%%

\section{The code\label{sec:code}}

An inversion code like the one presented in this work is an iterative numerical procedure that synthesises the Stokes profiles from different atmospheric models until the synthetic spectrum matches the observed one \citep[see, for instance,][]{delToroIniesta2016}. The process implies solving the RTE multiple times while perturbing the atmospheric parameters until an accurate solution is achieved -- in other words, until the difference between synthetic and observed profiles is as small as possible for the entire analysed spectral range. That accuracy (i.e. how small that difference is) is defined by the $\chi^2$ merit function:

\begin{equation}
\chi^2 = \frac{1}{N_f}\sum_{S=1}^{4}\sum_{i=1}^{n_{\lambda}}\left[ I_s^{\rm Obs}(\lambda_i)-I_s^{\rm Syn}(\lambda_i)       \right]^2 w_s(\lambda_i)^2
,\end{equation}where index $S$ runs over the four Stokes parameters, $i$ covers the wavelength samples up to $n_\lambda$, and $N_f$ stands for the number of degrees of freedom, that is, the difference between the number of observables (four Stokes profiles $\times$ $n_\lambda$) and the number of free parameters used in the inversion. As we will see later on, the latter corresponds to the sum of the number of nodes for each atmospheric parameter perturbed during the inversion. The weights $w_s(\lambda_i)$ are a regularisation term used to add more emphasis to a specific Stokes profile, for example Stokes~$I$, or a specific spectral line, when
fitting multiple spectral lines simultaneously, promoting better fits of those individual features. Hence, the weights are Stokes and wavelength dependent.

In the following, we delve into the terms introduced in the previous paragraphs so the future user of the code can better understand their numerical implementation in DeSIRe as well as their physical meaning.

\subsection{Forward modelling module}
DeSIRe combines the SIR and RH codes seamlessly and works with LTE
and NLTE lines and any combination of them. When DeSIRe inverts
NLTE lines, RH provides all the necessary elements for performing the
synthesis: the code runs the RH module for solving the
statistical equilibrium equations determining the required atomic
level populations and the forward solution of the RTE. The atomic
populations are used to calculate the NLTE departure coefficients
$\beta_{low,up}$ of the lower and upper levels involved
in the transitions producing the spectral lines that are used for the
inversion. The departure coefficients are needed for evaluating the
RFs (see Sect.\ \ref{RFsec}). DeSIRe outputs both the LTE and
NLTE spectral profiles as evaluated by RH when it runs in synthesis
mode. Minor differences in the transfer solution from the standard RH
code arise because DeSIRe interpolates the spectra to a uniform
wavelength grid. In addition, the spectrum is normalised to the local
continuum provided by the reference atmosphere the Harvard-Smithsonian reference atmosphere \citep[HSRA, ][]{Gingerich1971}.

\subsection{Response functions}\label{RFsec}
The RFs of the Stokes parameters
\citep{Landi1977} describe the variations induced in the Stokes
profiles by perturbations of an atmospheric quantity at a given
optical depth interval. They can be regarded as a sort of partial
derivative of the emerging Stokes parameters with respect to the
atmosphere's physical parameters.
%BRC
If we introduce a perturbation $\delta x_i$ of a
physical quantity $x_i$ at an optical depth interval
$(\tau_i,\tau_i+\Delta\tau)$, with $i$ running from
$i=1,n_{\tau}$ and introduce first-order perturbations in the RTE
(Eq. \ref{RTEpol}), the formal solution for the perturbed Stokes vector, $\delta
\vec{I}$, will be
\begin{equation}
  \delta \vec{I(0)}=\int_0^\infty \mbox{O}(0,\tau)
  \bf{\hat{\mathscr{S}}}(\tau) d\tau,
\label{pert1}
\end{equation}
with the perturbed source function
$\bf{\hat{\mathscr{S}}}(\tau)=\delta \mathscr{S} -
\delta\mbox{K}\vec{I}$. Writing
\begin {align}
\delta \mbox{K}=\frac{\partial\mbox{K}}{\partial x_i} \delta x_i\\
\delta \mathscr{S} =\frac{\partial \mathscr{S}}{\partial x_i} \delta x_i,
\label{pert2}
\end{align}
we can define the RF, $\vec{R}(\tau_i)$, of the Stokes parameters to
perturbations of a physical quantity $x_i=x(\tau_i)$ at an optical
depth $\tau_i$ \citep[see, for instance,][]{RuizCobo1994} as
\begin{equation}
  \vec{R}(\tau_i) \equiv \mbox{O}(0,\tau) \left[
    \frac{\partial \mathscr{S}}{\partial x_i} -
    \frac {\partial \mbox{K}}{\partial x_i} \vec{I} \right ],
   \label{RFeq}
\end{equation}
where the evolution operator $\mbox{O}(\tau,\tau')$ is the matrix
multiplying $\vec{I(\tau')}$ to obtain $\vec{I(\tau)}$ taking into
account only absorption processes:
\begin{equation}
  \frac {d \mbox{O}(\tau,\tau')}{d \tau'}= \mbox{K}(\tau)
  \mbox{O}(\tau,\tau').
   \label{Oeq}
\end{equation}
From Eqs.\ (\ref{pert1}) and (\ref{pert2}) we get that the
perturbation of the Stokes profile can be expressed as
\begin{equation}
\delta \vec{I(0)}=\sum_{i=1}^{n_{\tau}}\vec{R}(\tau_i) {\delta{x_i}} \Delta\tau.
\label{pert2}
\end{equation}
This is equivalent to defining the RF as the partial derivative of
the Stokes vector to changes in a physical quantity once it is
discretised:
\begin{equation}
\vec{R}(\tau_i)=\frac{\partial\vec{I}}{\partial{x_i}}.
\end{equation}
The evolution operator $\mbox{O}$ is obtained during the numerical
solution of the Stokes profile, so, we only need to evaluate the
derivatives of the absorption matrix $\mbox{K}$ and the source
function $\vec{S}$ to get the RF by using Eq. (\ref{RFeq}).

Inversion codes based on RFs have been generally used since the early
nineties, when SIR was released. SIR implements the LTE RFs
analytically and evaluates them while solving the RTE. As explained in
the introduction, DeSIRe makes use of the LTE RFs enhanced with the
FDC approximation \citep[see][]{SocasNavarro1998}, to provide a
suitable approximation of the full NLTE RFs.
  
The FDC approximation consists of neglecting the derivatives of the
departure coefficients $\beta^b_{low,up}$ (and the emission profile
$\psi^b$ in the case of partial frequency redistribution, PRD) with
respect to any of the model physical quantities when evaluating the
derivatives of $\mbox{K}$ and $S_I$ that appear in
Eq.~\ref{RFeq}. All the terms in this equation are estimated
using the SIR code assuming LTE relations, except for the departure
coefficients (and the emission profile $\psi^b$ if necessary), which
are calculated with the RH code. We wish to emphasise that the final
solution of the iterative process of the inversions consistently takes
into account the full complexity of the combined NLTE RTE and the statistical equilibrium equations, regardless of the
approximation of FDCs in the RFs, which only map out how to get to the final solution.

\begin{figure}
\begin{center} 
   \includegraphics[trim=0 0 0 0,width=8.5cm]{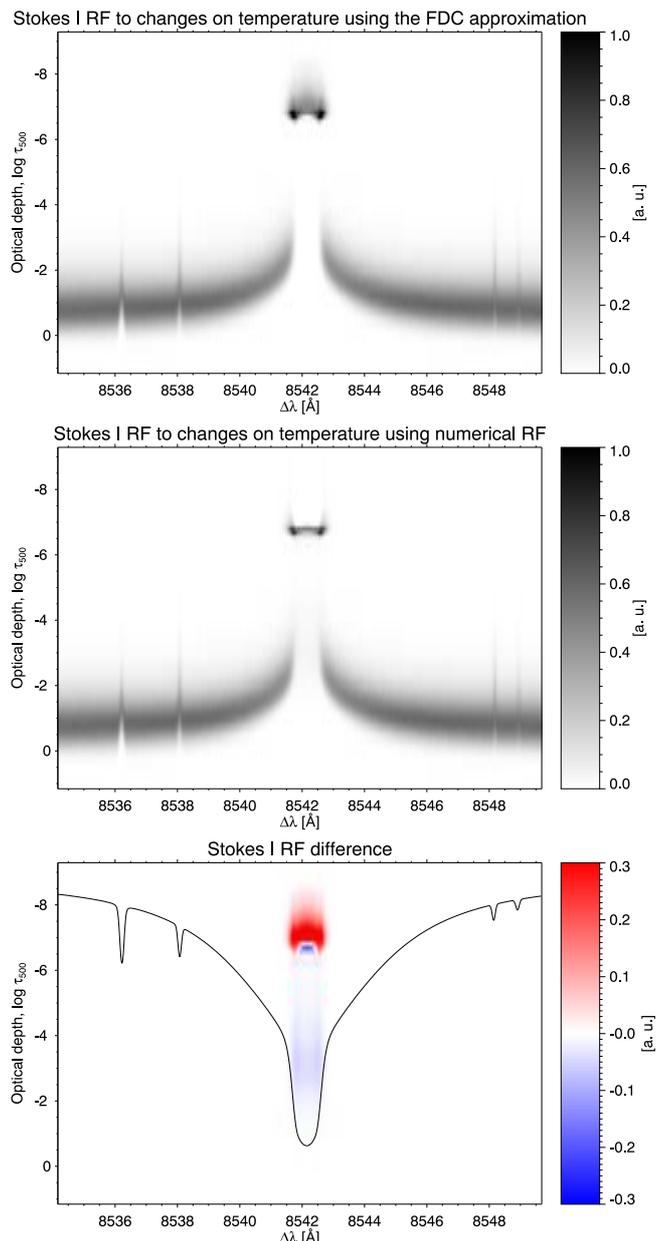}
 \vspace{-0.15cm}
 \caption{ \ion{Ca}{ii} 8542~\AA\ Stokes I RF to changes in
   temperature. The top panel shows the RF computed using the FDC scheme
   implemented in DeSIRe, the middle panel shows the RF computed using the
   numerically evaluated NLTE RF, and the bottom panel displays the
   difference (normalised to the maximum of the numerical RF) between
   both computations. The intensity profile of the \ion{Ca}{ii} 8542
   line is added in the bottom panel.}
 \label{FDC_compar}
 \end{center}
\end{figure}

To illustrate essence of the FDC approximation, we show in
Fig.\ \ref{FDC_compar} a comparison between
the FDC and numerically evaluated RF to changes in temperature for
Stokes $I$ in the \ion{Ca}{ii} 8542~\AA\ transition.  The top panel
shows the FDC RF used in DeSIRe. White colours indicate a lack of
sensitivity to temperature changes, while dark areas highlight the
spectral regions and heights where the intensity in the transition is
sensitive to those perturbations. The middle panel
displays the same quantity, obtained through a numerical
differentiation, which is more accurate, as it properly includes all
contributions in Eq.\ \ref{pert2}. The bottom panel shows the
difference between the two RFs, normalised to the maximum value of the
NLTE RF (middle). In the last panel we also plot the intensity profile
of the \ion{Ca}{ii} transition. It is clear that differences are
negligible outside the line's core wavelengths,
and that they increase closer to
the centre of the spectral line. In this case, we note deviations
of up to around 30\% from the NLTE numerical RF. Still, as we mentioned in
the previous paragraph, these differences imply that the code may
need to perform more iterations to reach the best solution, but that
this does not hinder the code from attaining it.
 
\subsection{Cycles, nodes, and equivalent response functions}\label{ERF}

From Sect. \ref{sec:RTE} we know that we need to specify the value
of temperature $T$, electron pressure $Pe$, microturbulence $\xi$,
line of sight (LOS) velocity $V_{LoS}$, and magnetic field vector ($B,
\gamma, \chi$) at every reference optical depth $\tau_{500}$ , to
evaluate the emerging Stokes profiles and the RFs.  In addition, to
reproduce some observational features, we also included three
additional uni-evaluated parameters that are not defined at every
optical depth but are assumed constant throughout the atmosphere: the macroturbulence amplitude, $Vmac$, the filling factor, $\alpha_f$, and
the stray light factor, $\alpha_{str}$.  We call this table of
physical quantities our model atmosphere and denote it as $x_i=
x(\tau_i)$, where $x$ stands for any physical quantity, $T$, $B$,
and so on. We note that we do not need to specify the gas pressure $Pg$, nor
gas density $\rho$ since both are determined from $T$ and $Pe$ through
the equation of state. In addition, geometrical height $z$ can be
derived by integration of Eq.\ \ref{opticaldepth}.

The vertical domain of a typical atmospheric model ranges from
log($\tau_{500})\sim [1,-8]$ with a step size of $\sim 0.1$. Hence, a
model atmosphere will contain around one hundred points in depth. In
general, spectral lines will not be sensitive to fluctuations spanning
a few steps in optical depth because the mean free path of a photon is
larger than log($\tau_{500})\sim 0.1$. Consequently, a perturbation in
any physical quantity with a very high spatial frequency will be
undetectable. The inversion code works iteratively to determine the
highest frequency fluctuation we should use, mimicking an expansion
series. We call a single instance of these
iterations an iterative cycle. Typically, a constant or linear
perturbation through the whole atmosphere is specified in the first cycle;
in the second one, a parabolic perturbation is added and further
refinements are applied in successive cycles until appropriate
convergence is reached.

We define as nodes those points in optical depth at which the
perturbation of each cycle is evaluated, and let $\delta
x_j=x(\tau_j)$, with $j=1,...,n_{nod}$, be those perturbations in
physical quantities evaluated at these nodes $j$. The perturbations in
the intermediate depth points will be obtained after the following
interpolation:
\begin{equation}
\delta x_i=\sum_{j=1}^{n_{nod}} a_{ij} \delta x_j,
\label{interp1}
\end{equation}
where $a_{ij}$ are the interpolation coefficients. In general these
are obtained through cubic spline interpolation.
         
We can introduce the $\delta x_i$ expression in Eq. \ref{pert2}:
\begin{equation}
  \delta \vec{I(0)}=\sum_{i=1}^{n_{\tau}} \vec{R}(\tau_i) \left(
         {\sum_{j=1}^{n_{nod}} a_{ij}} \delta x_j\right) \Delta\tau = 
          \sum_{j=1}^{n_{nod}} \vec{\hat{R}}(\tau_j) \delta {x_j} \Delta\tau,
\label{pert3}
\end{equation}
where $\vec{\hat{R}}(\tau_j)$ are the equivalent response functions
(ERFs) at the nodes:
\begin{equation}
\vec{\hat{R}}(\tau_j) \equiv \sum_{i=1}^{n_{\tau}}  a_{ij}  \vec{R}(\tau_i).
\label{ERF}
\end{equation}

These ERFs are the ones DeSIRe uses during an inversion process (i.e. the code only computes the RF and solves the RTE on the optical
depth points where we have defined a node). For the remaining optical
depths DeSIRe interpolates the solution (i.e. the atmospheric parameters)
between these nodes.

\subsection{Inversion based on response functions}
DeSIRe combines the capabilities of SIR and RH  to solve the inversion of
the Stokes parameters in NLTE in a fast and flexible way. The flow
chart of the inversion module consists of the following six steps.

First, the code reads the observed Stokes profiles $I^{obs}$; the
  initial guess model atmosphere $a^j_i(\tau)$, where $i$ is the
  iteration index (0 in the case of the `initial guess' model); $j$ runs for
  every physical quantity for one or two model atmospheres
  (temperature, electronic pressure, magnetic field vector, LOS velocity, micro- and macro-turbulence velocities, filling
  factor and stray light contamination factor); the atomic models; and
  all the information required to synthesise the Stokes profile
  vector.
  
  Second, using RH's subroutines, DeSIRe synthesises
  $I_i^{syn}=I_i^\mathrm{NLTE}$ in NLTE computing the departure
  coefficients $\beta_{up}$ and $\beta_{low}$ from the atmospheric
  model $a_i$ for every transition line. The computation is done
  including multiple atoms and molecules, and considering complete or
  partial redistribution (CRD or PRD) for the transitions of
  interest. Using SIR subroutines, it synthesises $I_i^\mathrm{SIR}$
  in LTE and the approximated RFs (i.e. the LTE ones corrected by the
  FDCs and the ratio of the emission profile
  over the absorption profile, $\psi/\phi$) for every spectral line
  specified by the user.

Third, with these RFs, the SIR module makes an inversion iteration
  minimising the $\chi^2$ (i.e. the total squared difference between
  $I^{obs}$ and $I_i^{syn}$) and delivers a new atmospheric model,
  $a_{i+k}$, and the respective profile, $I_{i+k}^\mathrm{SIR}$ (with
  $k=1$ in this first step).

Fourth, if the maximum temperature difference between $a_{i+k}$ and
  $a_{i}$ is lower than a threshold specified by the user (we
  recommend a threshold of 10\% of the maximum value of the
  temperature), the code will start an LTE cycle, updating the RFs
  for every new model while keeping the departure
  coefficients constant. During this cycle the code minimises the
  total squared difference between $I^{obs}$ and
  $I_{i+k}^{syn}=I_i^\mathrm{NLTE}-I_i^\mathrm{SIR}+I_{i+k}^\mathrm{SIR}$. This
  iteration will run, increasing $k$ until the temperature
  perturbation exceeds the threshold, until $k$ reaches a given maximum,
  or until the $\chi^2$ reaches the desired target.

Fifth, when the temperature perturbation exceeds the given threshold
  for temperature changes, it goes back to step 2, increasing $i$ and
  re-evaluating the
  departure coefficients and $I_i^{syn}=I_i^\mathrm{NLTE}$ using RH.
  
Finally, once the code ends executing the previous points, it evaluates a
  final $I^{syn}=I^\mathrm{NLTE}$ running RH from the final atmospheric
  model $a^j(\tau)$.

Following these steps, the code can solve the NLTE inversion problem
minimising the total number of calls to the NLTE transfer problem in
order to reduce computational demands. The crucial element in the
DeSIRe flow chart is the temperature threshold. Should this threshold
be stringent (i.e. requiring an absolute change in
  temperature of e.g. less than 1\% between iterations), DeSIRe will run RH
internally a large number of times to
re-evaluate the FDCs. If it is set to a less stringent condition
(i.e. requiring that temperature changes by an absolute
  difference of less than 10-15\%), the number of calls to
RH is much reduced. Also, as an extra note, we want to clarify the meaning of
$I_i^\mathrm{SIR}$, used in the previous description. If DeSIRe runs
in LTE mode, for example if we are inverting only photospheric lines, then
that intensity is the one generated by SIR, with no DC correction. If
the code runs in NLTE, that intensity can correspond to two different
values. If we have started step 2, then that intensity corresponds to
the one generated by SIR, evaluated from the opacity and source function
corrected by the departure coefficients from
RH. If we are in step 4, that intensity corresponds to the one
generated by SIR from the atmosphere $a_{i+k}$ corrected with the DC
from the latest call to RH.

Integrating RH and SIR within the same code has particular
difficulties, the two most important of which are as follows. First, SIR and RH employ different opacity packages. However, DeSIRe
    always uses the NLTE profiles calculated by RH and its opacity
    package, whereas the only quantities calculated using the opacity
    package of SIR are the LTE RFs. 
Second, RH and SIR originally used different RTE solvers, while in
    DeSIRe the RH and SIR modules have been updated to work with the
    very same RTE solver (i.e. the DELO-Bezier formal solution
    presented in \citealt{delaCruzRodriguez2013}).

Finally, large datasets with many pixels can be inverted extremely efficiently
on massively parallel compute clusters by employing the Python wrapper
presented in \cite{Gafeira2021} with DeSIRe.

\begin{table*}
%\hspace{-0.5cm}
\normalsize
\begin{adjustbox}{width=0.960\textwidth}
  \bgroup
\def\arraystretch{1.25}
\begin{tabular}{lcccccccccccccccccc}
        \hline
Atom    & $\lambda$ [\AA] & $\log gf$ & $L_l$   & $U_l$  & $J_{low}$ & $J_{upp}$ &  $\sigma$ [a.u.] & $\alpha$ \\
        \hline
\ion{Fe}{i}   & 6301.5008   & -0.718   & $3d^6$(${}^5$D)$4s4p$(${}^3$P$^{\rm o}$) z ${}^5$P$^{\rm o}$ & $3d^6$(${}^5$D)$4s$(${}^6$D)5$s$ e ${}^5$D & 2 & 2 & 834 &  0.242 \\
\ion{Fe}{i}    & 6302.4932   & -1.131 & $3d^6$(${}^5$D)$4s4p$(${}^3$P$^{\rm o}$)     z ${}^5$P$^{\rm o}$ & $3d^6$(${}^5$D)$4s$(${}^6$D)5$s$ e ${}^5$D & 1 & 0 &  850 &  0.239 \\ 
        \hline
\ion{Fe}{i}    & 8496.990   & -0.821 & $3d^7$(${}^4$F)$4p$ y ${}^3$F$^{\rm o}$& $3d^7$(${}^4$F)$5s$ e ${}^3$F & 3 & 2 &  983 & 0.253 \\ 
\ion{Ca}{ii}   & 8498.023   & -1.312 & $3p^63d$          ${}^2$D         & $3p^64p$       ${}^2$P$^{\rm o}$ & 1.5 & 1.5 &  310 & 0.275 \\ 
\ion{Si}{i}    & 8501.544   & -1.530 & $3s^23p3d$ ${}^1$D$^{\rm o}$  & $3s^23p(^2$P$^{\rm o}_{3/2})4f$ $^2[5/2]$ & 2 & 2 &  1570 & 0.315 \\ 
\ion{Ni}{i}    & 8501.799   & -1.974 & $3d^9(^2$D$)4p$ ${}^1$F$^{\rm o}$ & $3d^9(^2$D$_{5/2})5s$ $^2$[5/2] & 3 & 2 &   790 & 0.234 \\
\ion{Si}{i}    &  8502.219   & -1.260 & $3s^23p3d$ ${}^1$D$^{\rm o}$ & $3s^23p(^2$P$^{\rm o}_{3/2})4f$ $^2[5/2]$ & 2 & 3 &   1570 & 0.315 \\
        \hline
\ion{Si}{i}   & 8536.164   & -0.910 & $3s^23p3d$ ${}^3$F$^{\rm o}$& $3s^23p(^2$P$^{\rm o}_{3/2})5f$ $^2$[7/2] & 2 & 3 &   2971 & 0.347 \\ 
\ion{Fe}{i}    & 8538.015   & -1.400 & $3d^6({}^5$D$)4s4p(^1$P$^{\rm o})$ x ${}^5$D$^{\rm o}$& $3d^6({}^5$D$)4s(^6$D$)4d$ e ${}^7$G & 4 & 4 &   786 & 0.269 \\ 
\ion{Ca}{ii}   & 8542.091   & -0.362 & $3p^63d$          ${}^2$D & $3p^64p$          ${}^2$P$^{\rm o}$ & 2.5 & 1.5 &  290 & 0.275 \\ 
\ion{Ti}{i}    & 8548.088   & -1.187 & $3d^3({}^2$G)$4s$ a ${}^3$G & $3d^2({}^1$D$)4s4p(^3$P$^{\rm o})$ x ${}^3$F$^{\rm o}$ & 3 & 2 &   267 & 0.252 \\
\ion{Cr}{i}    & 8548.851   & -2.066 & $3d^5({}^4$P)$4s$ a ${}^5$P& $3d^4({}^5$D$)4s4p(^3$P$^{\rm o})$ z ${}^5$D$^{\rm o}$ & 2 & 2 &   272 & 0.244 \\   
\ion{Si}{i}    & 8550.353   & -1.480 & $3s^23p4p$ ${}^1$D$^{\rm o}$ & $3s^23p(^2$P$^{\rm o}_{3/2})7s_{1/2}$ $(\frac{3}{2},\frac{1}{2})^{\rm o}$& 2 & 2 &   4763 & 0.225 \\  
        \hline
  \end{tabular}
  \egroup
\end{adjustbox}
\caption {Spectral lines included in the inversion test,
  representative of the DKIST/ViSP level 2 configuration proposal. The columns, from left to the right, show the atomic species, the line
  core wavelength, the $\log gf$ of the transition, the dominant
  configuration and term designation for the lower and upper levels,
  the total angular momentum quantum numbers of the lower and upper
  levels, and the calculated line broadening cross-section $\sigma$
  due to hydrogen atom impact at a collision velocity of $10^4$~m/s
  and corresponding velocity dependence parameter $\alpha$. Units for
  the wavelength are \AA\  and atomic units (they correspond to the
  square of the Bohr's radius, i.e. 2.80$\times 10^{-21}$ m$^2$) for
  the cross-section $\sigma$. The velocity parameter
  $\alpha$ is dimensionless.  } \label{lines}
\end{table*}

\section{Numerical experiments}\label{inver3D_sec}
\begin{figure}
\begin{center} 
   \includegraphics[trim=0 0 0 0,width=8.5cm]{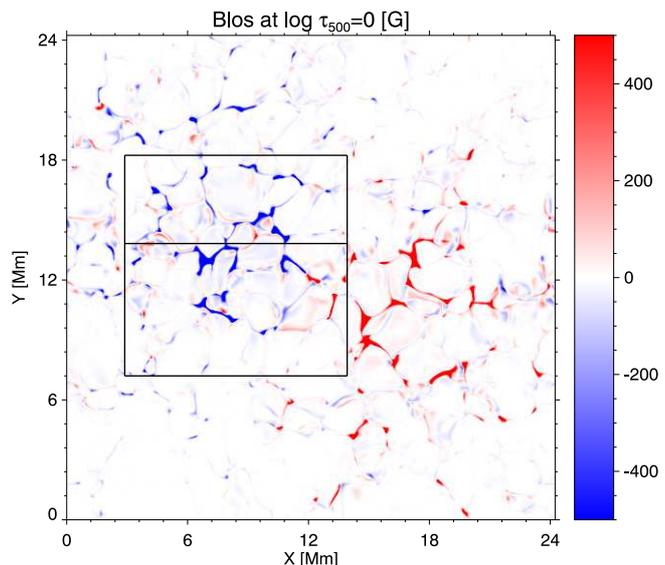}
 \vspace{-0.15cm}
 \caption{Snapshot 385 from the Bifrost enhanced network
   simulation. We show the longitudinal component of the magnetic
   field strength at log~$\tau_{500}=0$. We used two regions in this
   work, one within the squared box and the one highlighted with
   the horizontal line.}
 \label{Context}
 \end{center}
\end{figure}

During the development of DeSIRe, it was extensively tested with
different atmospheric models and conditions to evaluate its robustness
and accuracy. For clarity, we focus only on the tests performed with
3D realistic numerical simulations in this work since these allow us
to directly compare the inversion results with the original
atmosphere. We used snapshot 385 of the Enhanced Network simulation
described in \cite{Carlsson2016} and developed with the Bifrost code
\citep{Gudiksen2011}. The snapshot covers a surface of
$24\times24$~Mm$^2$ with a pixel size of 48~km, and a vertical domain
ranging from 2.4~Mm below to 14.4~Mm above the continuum average
optical depth unity at $\lambda=500$~nm. The simulated scenario,
albeit simpler than recent numerical simulations such as those presented
in \cite{Hansteen2017, Hansteen2019} and \cite{Cheung2019}, contains a
configuration with strong network patches and quiet areas (see
Fig.~\ref{Context}). The mentioned structures allow the
capabilities of the spectral lines inversions to be ascertained from quiet-Sun
spectropolarimetry. Moreover, the simulation is publicly
available\footnote{\url{http://sdc.uio.no/search/simulations}}, so the
present work can be independently verified and allows our inversion
results to be compared with those of different inversion codes.

We assumed disk centre observations (i.e. $\mu=1$, where
$\mu=\cos(\theta)$ and $\theta$ the heliocentric angle). We used the
abundance values of the different atomic species given in
\cite{Asplund2009}. We did not study the influence of noise or limited
spatial and spectral resolution, and we did not include any
microturbulence enhancement either. We are primarily interested in
examining the code's capabilities for inverting multiple lines in LTE
and NLTE simultaneously.

\subsection{Simulating DKIST level-2 ViSP observations}
We study the accuracy of DeSIRe when inverting a large field of view (FOV) of the
Bifrost simulation first (see the square in Fig.~\ref{Context}). The
synthesis is done with DeSIRe after transforming the simulation to an
optical depth grid. We focus in this section on the \ion{Fe}{i} 630~nm
line pair plus the two \ion{Ca}{ii} spectral windows presented
below. These spectral regions are selected based on the plans for
DKIST's Visible Spectro-Polarimeter \citep[ViSP,][]{deWijn2012,2021AJ....161...89D} level-2 standard
inversion products\footnote{\url{https://nso.edu/ncsp/}}. In addition,
these spectral lines have been selected for spectropolarimetry with
ViSP during the DKIST Cycle 1 Proposal Call. They will also be
observed (in at least the red part of the spectrum) by the upcoming
Sunrise Chromospheric Infrared Polarimeter
\citep[SCIP;][]{Katsukawa2020} instrument that will operate on board the Sunrise
\citep{Solanki2010, Barthol2011} balloon's third flight, programmed for
2022.  Thus, we believe it is an excellent opportunity to test the
code in this section focusing on their target spectrum. The atomic
information of the strongest spectral lines in the selected spectral
windows is shown in Table~\ref{lines}. The information was obtained
from either the National Institute of Standards and Technology
\citep[NIST;][]{nist2018} or the R.~Kurucz \citep{Kurucz1995}
databases. The broadening of the spectral lines by collisions with
neutral hydrogen atoms is computed using the Anstee, Barklem, and
O'Mara (ABO) theory \citep[e.g.][]{Anstee1995, Barklem1997},
specifically the abo-cross calculator code \citep{Barklem2015}, which
interpolates in pre-computed tables of line broadening parameters. A
description of how to use the code can be found in \cite{Barklem1998}.

\subsection{Inversion configuration}
We inverted the synthetic profiles computed with DeSIRe, starting from five
possible initial guess atmospheres derived by applying linear
perturbations to the FAL-C \citep{Fontenla1993} atmosphere and using the
set of nodes presented in Table~\ref{nodes}. We pick the best fit from
the converged solutions with five different initialisations.
\begin{table}
%\hspace{-0.5cm}
\normalsize
\begin{adjustbox}{width=0.47\textwidth}
  \bgroup
\def\arraystretch{1.25}
\begin{tabular}{l|cccccccccccccccccc}
        \hline
\multirow{2}{*}{Parameter}     & & \multicolumn{2}{c}{Nodes}   \\
     & Cycle 1 & Cycle 2 &  Cycle 3 & Cycle 4  \\
        \hline
Temperature    & 2 & 5 &  6 & 9 \\
Microturbulence    & 1 & 1 &  2 & 5 \\  
LOS velocity    & 1 & 2 &  5 & 6 \\     
Magnetic field    & 1 & 2 &  5 & 6 \\
Inclination    & 1 & 2 &  5 & 6 \\      
Azimuth    & 1 & 1 &  2 & 2 \\
        \hline  
  \end{tabular}
  \egroup
\end{adjustbox}
\caption {Nodes used on the inversions presented in Sect.~\ref{inver3D_sec}.
} \label{nodes}     
\end{table}

%\subsubsection{Two Tests}
%
We performed two different tests using the previous
configuration. First, we synthesise what we called ViSP level 2-like
data where the spectral lines presented in Table~\ref{lines} are used,
solving the \ion{Ca}{ii} populations in NLTE. This inversion is done
over the squared FOV highlighted in Fig.~\ref{Context}. In the
second case, we performed the inversion of the pixels highlighted by the
horizontal line enclosed by the square. In that case, we invert
multiple atoms in NLTE to test the code capabilities and stability
under these conditions.
\begin{figure*}
\begin{center} 
   \includegraphics[trim=0 0 0 0,width=16.0cm]{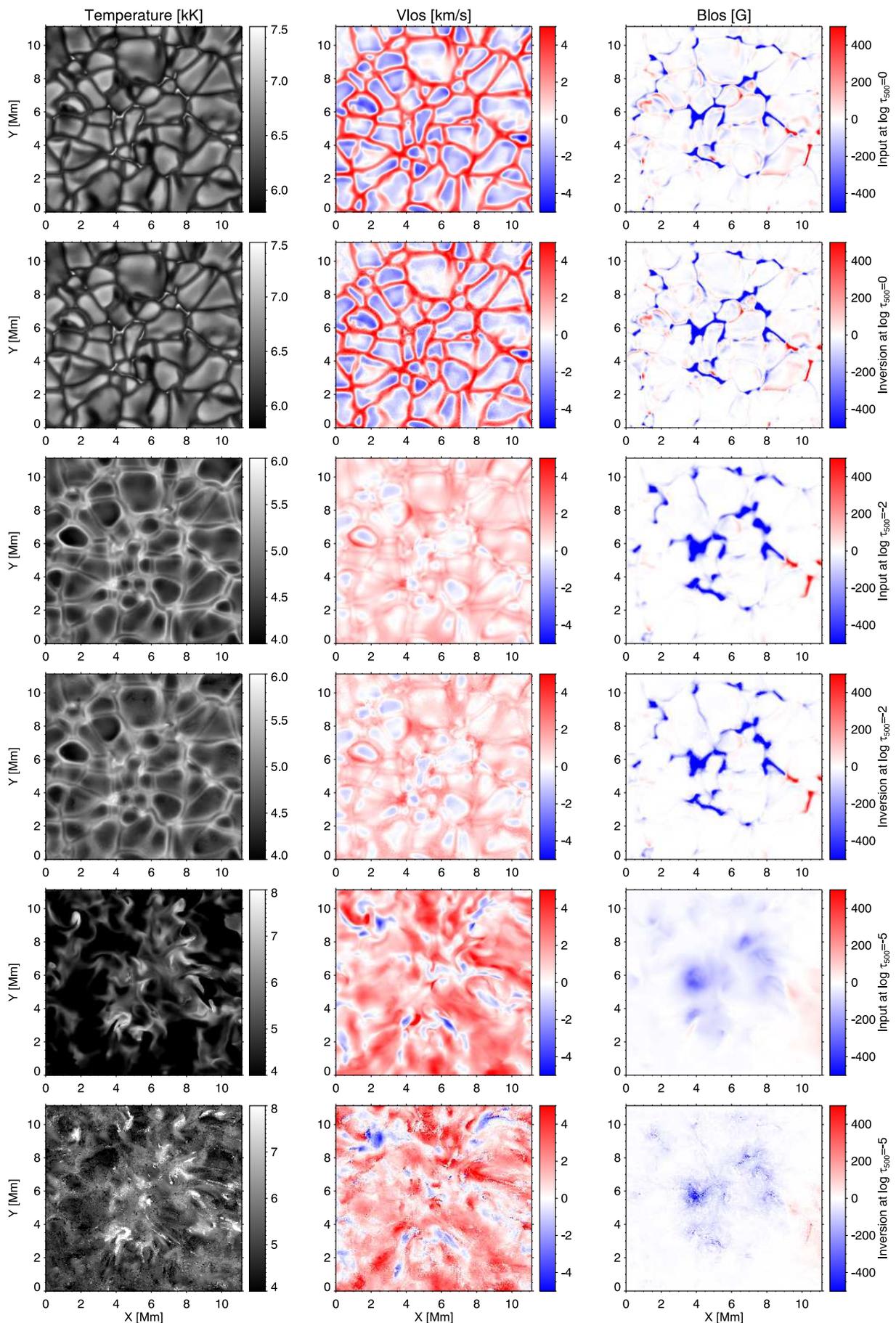}
 \vspace{-0.15cm}
 \caption{Comparison between the input atmosphere (odd rows) and the
   one inferred with DeSIRe (even rows). From left to right, we show
   the temperature, the LOS velocity, and the magnetic field. From top to
   bottom, we display the atmospheric parameters at three reference
   layers at log~$\tau_{500}$=[0,-2,-5]. The FOV corresponds to the
   spatial domain within the squared region in Fig.~\ref{Context}.}
 \label{Inver3D}
 \end{center}
\end{figure*}
\begin{figure*}
\begin{center} 
   \includegraphics[trim=0 0 0 0,width=18.0cm]{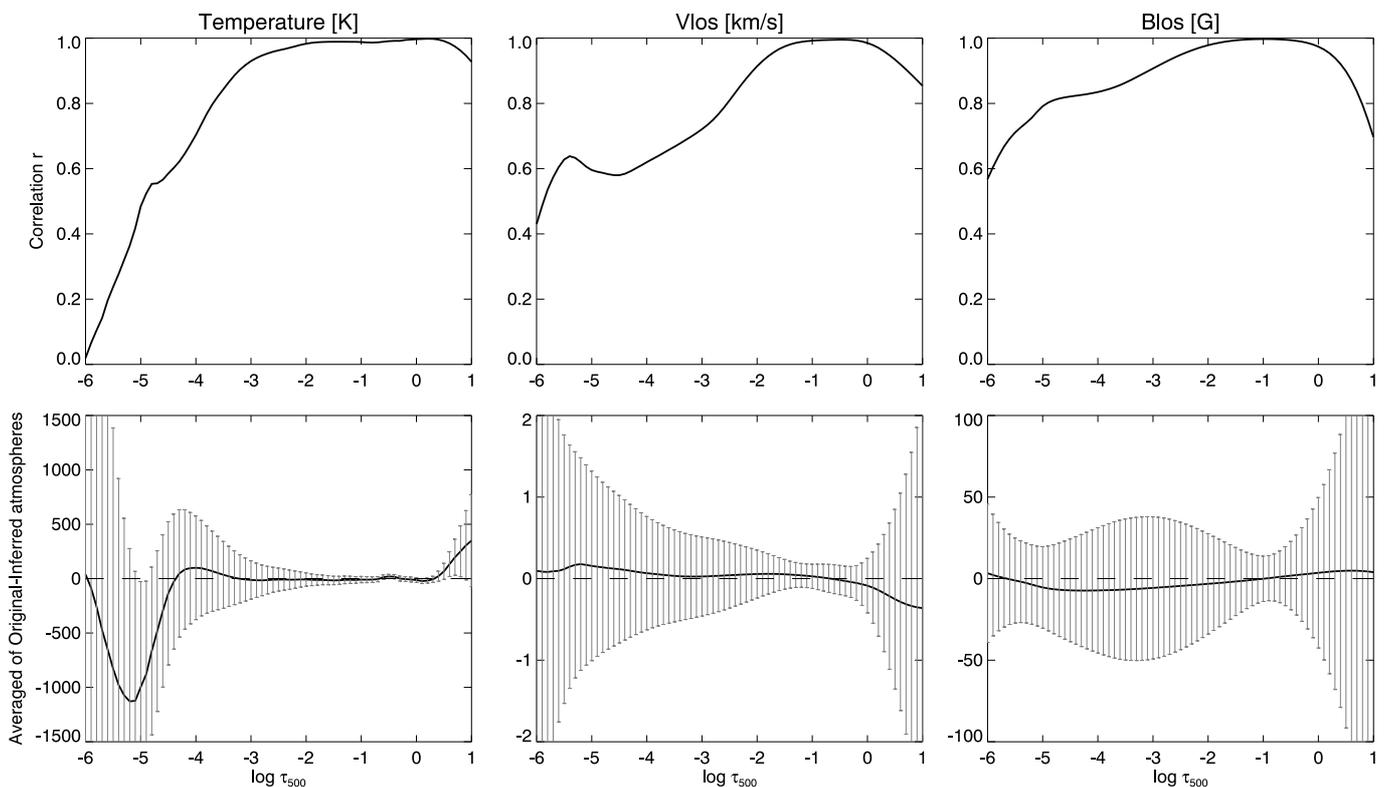}
 \vspace{-0.15cm}
 \caption{Accuracy of the inversion code for inferring complex atmospheres. The top row displays the correlation between the input
   and the inferred atmospheric parameters presented in
   Fig.~\ref{Inver3D}. The bottom row shows the average value of the
   difference between the input and the inferred atmosphere over the
   entire inverted FOV. Error bars designate the standard deviation of the
   difference. From left to
   right, columns correspond to the temperature, LOS velocity, and LOS
   magnetic field. Each panel displays the variation in the
   correlation and differences with optical depth.}
 \label{Inver3D_correl}
 \end{center}
\end{figure*}
\begin{figure*}
\begin{center} 
   \includegraphics[trim=0 0 0 0,width=18.0cm]{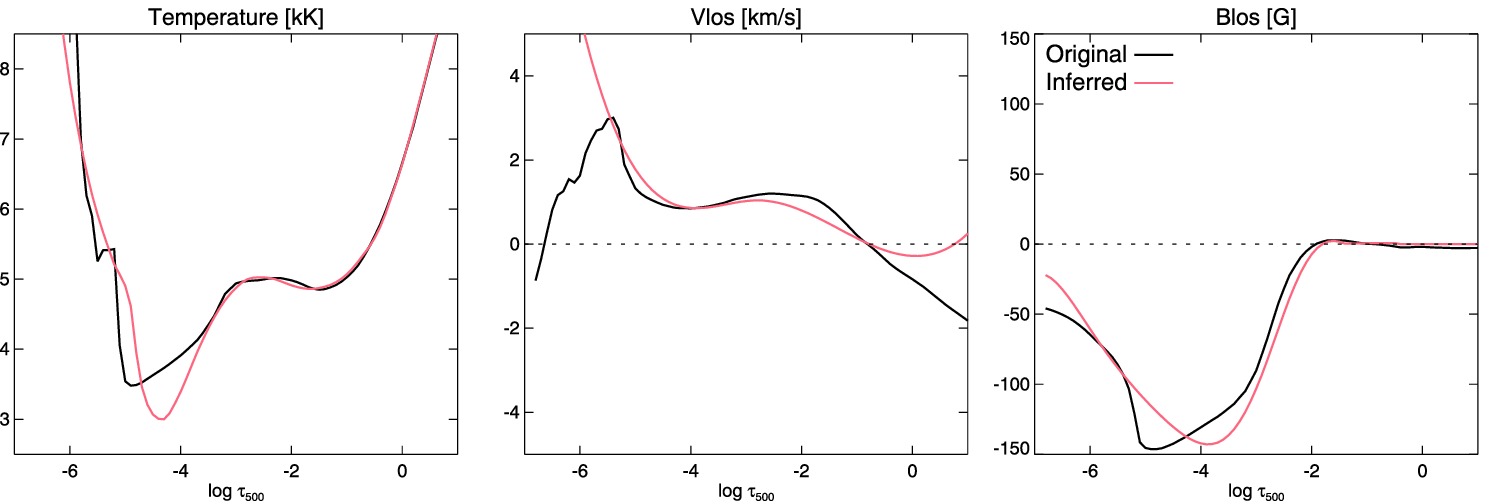}
 \vspace{-0.15cm}
 \caption{Stratification of temperature, LOS velocity,
   and LOS magnetic field (from left to right) in one pixel. Black
   indicates the original physical parameters used during the
   synthesis, and red designates the inferred atmospheres using the
   configuration explained in Sect.~\ref{inver3D_sec}. The pixel is
   located at [6.2,12.6]~Mm in Fig.~\ref{Context}.}
 \label{Inver3D_atmos}
 \end{center}
\end{figure*}

\subsection{DKIST level-2 ViSP configuration inversion test}
The results of the inversion of the square subfield in
Fig.\ \ref{Context} are shown in Fig.~\ref{Inver3D}, where
quantities in odd rows correspond to the original atmosphere at
log~$\tau_{500}=[0,-2,-5]$\footnote{We always refer to the optical
depth at 500~nm when using log~$\tau_{500}$.} while even rows
display the inferred atmospheric quantities at the same optical
depths. We can see that, in general, the resemblance between the
original and inferred temperature, LOS velocity, and LOS magnetic
field is very good. In particular, the values are almost identical in
lower layers. In the case of upper layers, we see more
differences. The code cannot accurately retrieve the temperature in
the cool areas of the original atmosphere (first column, $\tau_{500}
=-5$), although it is more accurate in reproducing the velocity and
magnetic field. Thus, taking into account that we are starting from a
FALC atmosphere with the LOS velocity and the magnetic field vector
constant with height, we believe the code is behaving well and is
retrieving a solution that is close to the original one (despite the
finite number of nodes used in the process).

We note that the averaged inversion time per pixel is 5 minutes on a
standard 2.1 GHz type CPU, i.e. Xeon Gold 6152. So, the code is capable of inverting
the whole $230\times 230$ pixel subfield in a little over 3 hours on a
1000 CPU cluster, showing that the speed of the DeSIRe code is
suitable for inverting large DKIST FOVs in a reasonable time.

\begin{figure*}
\begin{center} 
   \includegraphics[trim=0 0 0 0,width=18.0cm]{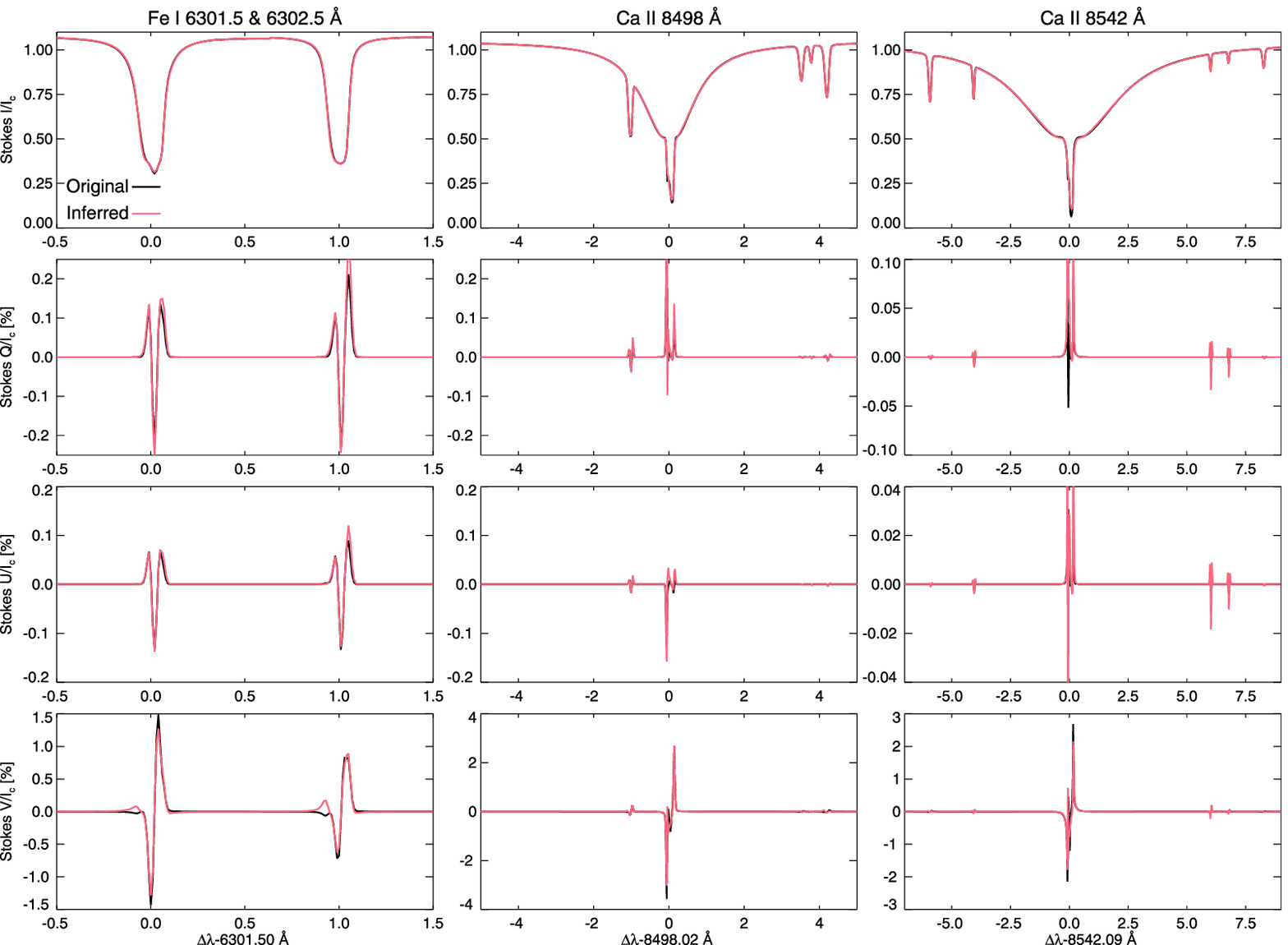}
 \vspace{-0.15cm}
 \caption{Stokes ($I$, $Q$, $U$, $V$)
   parameters (from top to bottom) for the \ion{Fe}{i} 6301.5 $\&$ 6302.5~\AA \ (left),
   \ion{Ca}{ii} 8498~\AA \ (middle), \ion{and Ca}{ii} 8542~\AA \ (right)
   spectral lines. Black designates the original Stokes profiles, and
   red corresponds to the inferred spectra using the configuration
   explained in Sect.~\ref{inver3D_sec}. The pixel is located at
   [6.2,12.6]~Mm in Fig.~\ref{Context}, and the associated atmosphere
   is presented in Fig.~\ref{Inver3D_atmos}.}
 \label{Inver3D_profiles}
 \end{center}
\end{figure*}

To illustrate the good correspondence between original atmospheric
quantities and the results of the inversion, we show their correlation as a
function of height in Fig.~\ref{Inver3D_correl}. We note that we are
matching on average the input atmosphere in lower layers, but that the
correlation drops in higher layers, above log~$\tau_{500}=-3$. This
reduction of the correlation can be due to several factors; the
assumptions in the code itself, the inversion configuration, or the
limited sensitivity of the employed spectral lines to properties of
the different layers. In particular, the LTE lines are sensitive below
log~$\tau_{500}\approx-2.5$, while the two \ion{Ca}{ii} infrared lines
are most sensitive at layers deeper than log~$\tau_{500}\approx-5.0$
\citep{QuinteroNoda2017} so there is a region in between where the
spectral lines we employ in this setup are less
sensitive. Accordingly, we can see a slight decrease in the
correlation for all the atmospheric parameters between
log~$\tau_{500}\approx[-3,-5]$. We believe the drop is due to the gap
in the sensitivity between the iron lines and the calcium lines in the
infrared. We complement the previous analysis by plotting the
  average values for each optical depth of the difference between the
  input and the inferred atmosphere in the bottom row of the same
  figure. We add an error bar that corresponds to the standard
  deviation of the same difference over the entire FOV for each
  optical depth. The results are in agreement with what we explained
  above, with the differences close to zero in general between
  log~$\tau_{500}\approx[0,-5]$. Only the temperature shows larger
  discrepancies above log~$\tau_{500}\approx-4.5$. We believe those
  more significant differences correspond to the cool areas on the
  input atmosphere that our inversion run could not match (see
  Fig.~\ref{Inver3D}).

\begin{figure}
\begin{center} 
   \includegraphics[trim=0 0 0
     0,width=8.5cm]{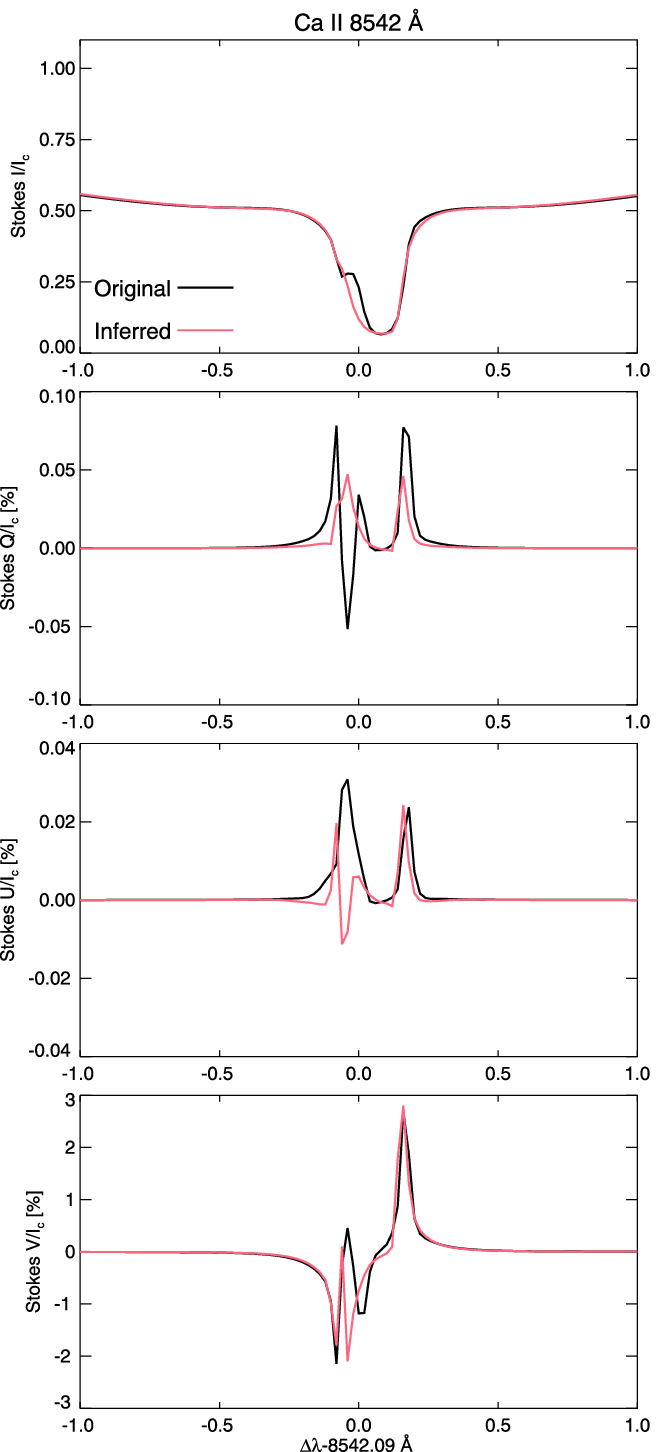}
 \vspace{-0.15cm}
 \caption{Zoomed-in view of the \ion{Ca}{ii} 8542~\AA \ Stokes
   profiles presented in Fig.~\ref{Inver3D_profiles}. The format is
   the same.}
 \label{Inver3D_profiles_zoom}
 \end{center}
\end{figure}

To show the code's capability to achieve close fits to all employed
LTE and NLTE spectral lines, we examined one pixel in detail. We chose
a pixel where the magnetic field stratification is complex, that is, the
magnetic field is weak at lower layers and gets stronger at higher
heights. This pixel seems to be representative of the canopy
effect. It is located at [6.2,12.6]~Mm in Fig.~\ref{Context}, and it
also has a temperature and LOS velocity stratification that vary on a
small spatial scale. Thus, we believe it is a good benchmark of the
code behaviour and the configuration used in this test. In
Fig.~\ref{Inver3D_atmos}, we show the original values of
temperature, LOS velocity and LOS magnetic field (in
black) and their recovered stratification (in red) for the selected
pixel. We note that the inferred temperature traces with high accuracy
the original atmosphere in most of the layers. There are slight
differences at around log~$\tau_{500}\approx-4.3$, but in general, the
resemblance is very close. The same can be said for the LOS velocity,
where the inferred values match the original ones up to
log~$\tau_{500}\approx-5.5$ before it deviates, probably due to the
lack of sensitivity at the top of the atmosphere. The same behaviour
is found in the LOS magnetic field, where the code tracks the original
atmosphere at all heights (although with less accuracy at
log~$\tau_{500}\approx-4.7$). Thus, we confirm that the code is
working correctly and recovers the input
atmosphere with high accuracy. Since we can directly compare inverted values with the
original atmospheric quantities, we do not show any estimates of the
uncertainty in these graphs. However, in the future, when processing
actual observations, the uncertainty of the inversion results can, for
instance, be computed from the maximum amplitude of the total RF (all
spectral lines combined) at each optical depth.

We display in Fig.~\ref{Inver3D_profiles} the comparison between the
original forward calculated Stokes parameters (in black) of the
selected pixel with the values from the inversion (in red) in all
three spectral bands, centred on the \ion{Fe}{i}, \ion{Ca}{ii}
8498 \AA,\ and \ion{Ca}{ii} 8542 \AA\ lines, respectively. The code
fits the Stokes $I$ and $V$ profiles very accurately in all three
bands (including the overlapping photospheric line in the calcium
triplet bands, apart from perhaps the $V$ profile of the 8542
\AA\ line), explaining why the temperature and LOS velocity and
magnetic field are reproduced very well in all layers. The amplitudes
of the linear polarisation signals, $Q$ and $U$ are slightly less well
produced, indicating that the inferred orientation of the magnetic
field would be less accurately recovered. We further explore
  the accuracy of the fits of the chromospheric lines, in particular
  the \ion{Ca}{ii} 8542 \AA\, in
  Fig.~\ref{Inver3D_profiles_zoom}. There, we zoom in on the
  spectral range analysed in Fig.~\ref{Inver3D_profiles} to
  understand better the similarities between the original forward
  calculated Stokes profiles (black) and those inferred from the
  inversion (red). In general, we can see that, as mentioned above,
  the code is obtaining similar profiles to the original ones, even
  being complex ones with multiple lobes with different
  amplitude. Although it is true that the fit of the linear
  polarisation signals is worse than that of Stokes~$I$ and $V$.

Therefore, the code can fit the diverse set of spectral lines very
well at all wavelengths, indicating that this setup is very well
suited to simultaneously recovering atmospheric properties in both the
photosphere and chromosphere. It is thus very well suited to analyse
data from the DKIST/ViSP instrument (with its three arms configured to
observe all three spectral bands employed here) and the Sunrise/SCIP
instrument observing the 8498 and 8542~\AA\ bands
\citep{Katsukawa2020}.

\subsection{Multi-atom inversion test}\label{Multisec}

The strategy for synthesis and inversion in this section is
the same as before. However, we changed the spectral lines of
interest. We only synthesised two photospheric lines in LTE: the
\ion{Fe}{i} ones at 6301.5 and 6302.5~\AA\, and we combine them with
multiple NLTE transitions to test the accuracy and speed of the
code. We compute the \ion{Mg}{i} $b$ lines, the \ion{Na}{i} and
\ion{K}{i} D$_1$ $\&$ D$_2$ lines, and the \ion{Ca}{ii} 8498 and
8542~\AA \ lines in NLTE. For the Mg spectral lines, we use the atom
model presented in \cite{QuinteroNoda2018}. We employ the atom models
included with RH's source code distribution for the rest of the
transitions.

\begin{figure*}
\begin{center} 
 \includegraphics[trim=0 0 0
   0,width=18.0cm]{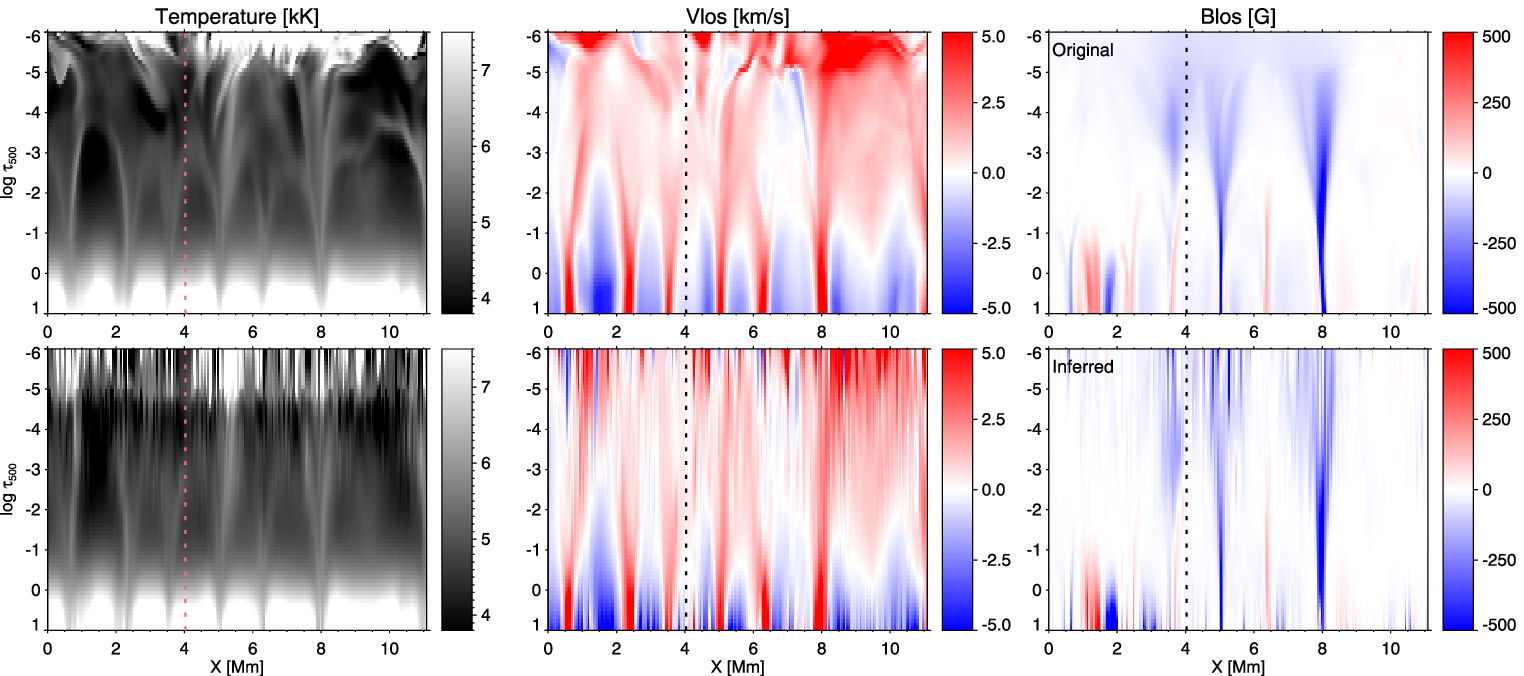}
 \vspace{-0.15cm}
 \caption{Comparison between the input atmosphere (top) and the atmosphere
   inferred with DeSIRe (bottom) for the inversion test described in
   Sect.~\ref{Multisec}. From left to right, we show the temperature,
   the LOS velocity, and the magnetic field. The FOV corresponds to the
   horizontal line enclosed within the squared region in
   Fig.~\ref{Context}. Abscissae correspond to the X axis in
   Fig.~\ref{Context}, and ordinates show the evolution with the
   height of the atmospheric parameters.}
 \label{Inver2D_Multi}
 \end{center}
\end{figure*}

\begin{figure*}
\begin{center} 
   \includegraphics[trim=0 0 0
     0,width=18.0cm]{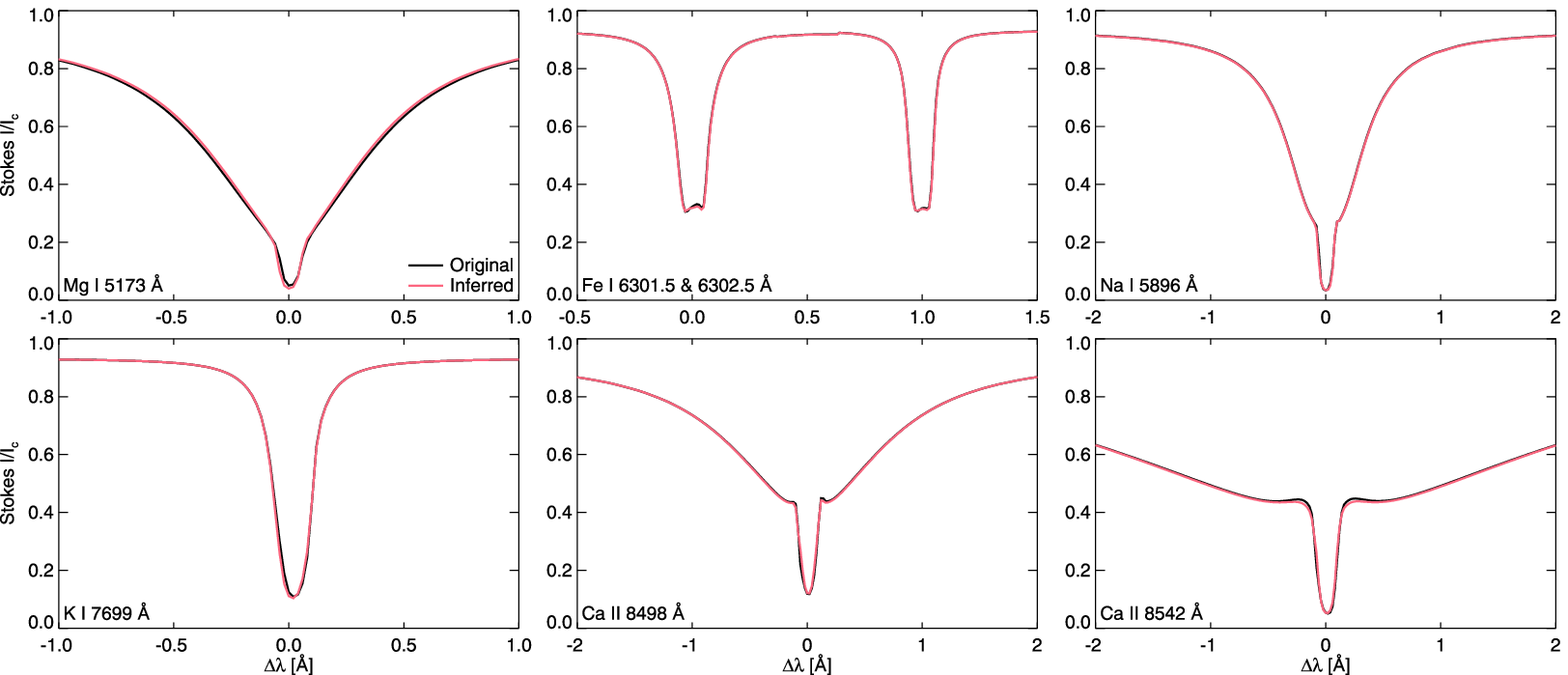}
 \vspace{-0.15cm}
 \caption{Results from the multi-atom inversion test described in
   Sect.~\ref{Multisec}. Panels show the intensity, with black the
   input profile and red the inferred profile from DeSIRe. The pixel is
   located at X=4~Mm in Fig.~\ref{Context} (see dotted line).}
 \label{Inver2D_Multi_profiles}
 \end{center}
\end{figure*}

We present the multi-atom inversion results in
Fig.~\ref{Inver2D_Multi} using a vertical cut through the atmosphere
(corresponding to the horizontal line in Fig.\ \ref{Context}, with the
depth scale in log~$\tau_{500}$). We aim to examine the height
stratification of the atmospheric parameters using multiple NLTE
transitions with various heights of formation. Moreover, we want to
evaluate the code's capability for retrieving the smooth variation
from pixel to pixel. We remark here that the code works under the
1.5D approximation, inverting pixel by pixel, and we have not
included any regularisation between neighbouring pixels in the present
version. In other words, DeSIRe does not have any information or
pre-conditioning from adjacent pixels when inverting a given
pixel. Thus, our motivation is to determine whether the code can
recover the spatial smoothness of the original simulation.

As in the previous test, we can see in Fig.~\ref{Inver2D_Multi} that
the code is very reliable: the inferred atmospheric properties
(bottom) closely match the original ones (top). We notice the impact
of the discretisation of the inversion (see Table~\ref{nodes}), but
the height variation still resembles that of the original
atmosphere. We also detect that upper layers above
log~$\tau_{500}\approx-5.5$ are recovered considerably less
accurately, mainly in temperature. This results from the lack of
sensitivity of the spectral lines in these layers, more than a failing
of the code. Interestingly, we observe that complex magnetic field
configurations are recovered in most cases, even where the field
changes sign with height. Even if we do not add any regularisation in
this version of the code, we also find a smooth transition between
pixels. For instance, on the LOS velocity panels (middle), where the
changes from upflows, null velocity, and downflows are preserved with
high accuracy in the inferred atmospheres (taking into account that we
are starting the inversion with a constant LOS velocity and magnetic
field). Finally, we show in Fig.~\ref{Inver2D_Multi_profiles} an
example of the accuracy of DeSIRe when inverting multiple NLTE
spectral lines simultaneously. We only plot the intensity profile to
reduce the number of graphs. The fits are very accurate, indicating
that the code is stable even when combining such a variety of spectral
lines and atomic models, forming in diverse ways.

\section{Summary and conclusions\label{sec:discussion}}

This work presents a numerically highly efficient code for
inverting polarimetric data of lines that form under general NLTE
conditions. In particular, we describe how we combine the LTE
inversion code SIR with the NLTE forward solver RH through the
technique of FDCs (and fixed emission profiles
in the case of PRD lines) into our new code, DeSIRe. The SIR code employs
RFs that can be derived analytically under LTE
conditions. In DeSIRe, the RFs of NLTE lines are
evaluated by correcting the analytical RFs as
formulated for LTE conditions with the NLTE departure coefficients
from RH. In a series of iterative steps, an initial guess atmosphere
is corrected by alternately solving the general NLTE equation of
radiative transfer with RH, handing the evaluated spectrum, the
departure coefficients, and the emission profiles of relevant lines to
SIR, and evaluating the NLTE-corrected RFs. These are
then used in SIR to correct the guess atmosphere to improve the
spectral line fit. This process is iterated until convergence is
reached (i.e.
a sufficiently close match between observed and modelled spectra). DeSIRe can invert NLTE lines of different atomic species
simultaneously, solving the combined equations of radiative transfer
and statistical equilibrium consistently, including the effects of
partial frequency redistribution. Our experience is that the
implemented inversion process is stable and efficient. It is worth
noting that while the FDC method assumes that
the derivatives of the departure coefficients (and the emission
profile in the case of PRD) with respect to the atmospheric parameters
that need to be determined are negligible, this approximation is only
used for the convergence process. The final solution, once converged,
reflects the fully consistent simultaneous solution of the NLTE
equations of both radiative transfer and statistical equilibrium.

We describe several numerical tests in Sect.\ \ref{inver3D_sec} that we used to
validate the performance of the new code. Starting from a snapshot of
a Rad-MHD simulation of solar convection that includes chromospheric
physics (Fig. \ref{Context}), we calculated the emergent spectra in
three wavelength bands centred around the \ion{Fe}{i} 6300
\AA\ doublet and the \ion{Ca}{ii} 8498 and 8542 \AA\ IR triplet
lines, respectively. Then, starting from a series of five initial
atmospheres derived from the FALC standard quiet-Sun model, we inverted
all lines in the three bands and compared the stratification of the
derived physical parameters with the original values in each pixel. In
a second test, we performed the same type of exercise in a vertical slice
(see the horizontal marking line in Fig.\ \ref{Context}) of the same
snapshot, but using a different line set that includes spectral lines
from multiple atomic species (\ion{Ca}{ii}, \ion{Mg}{i}, \ion{Na}{i},
and \ion{K}{i)} in NLTE. We limited ourselves here to these numerical
experiments for verification purposes of accuracy and reliability,
deferring tests on actual data to a later paper.

In both test cases, we find that DeSIRe is able to
accurately fit the line profiles in all lines and all four Stokes
parameters (Figs.\ \ref{Inver3D_profiles} and
\ref{Inver2D_Multi_profiles}) and accordingly is able to recover the
physical parameters from the spectra reliably. In both cases, we find
that the most significant deviations between inverted and original
temperatures occur at the top of the atmosphere, for log$\tau_{500} <
-5$ (see Figs. \ref{Inver3D_correl} and \ref{Inver2D_Multi}), but
we ascribe that more to the physics of the formation of scattering lines in
the chromosphere, the emergent intensity of which has inherently
limited sensitivity to local thermodynamic conditions, than to a
failure of the implemented inversion mechanism.

In our test, we find that a typical NLTE inversion takes 2-5 min on a
modern 2.1 GHz CPU, making the inversion process significantly faster
than for inversion techniques that employ numerical RFs, and making clear that we have achieved our goal of building
an efficient and robust inversion engine. The efficiency of the new
code, combined with the parallel Python wrap that enables the
distribution of the inversion process in individual, independent
pixels over the CPUs of massively parallel computers, provides us with
the promise that we can manage the inversion of large high-resolution
datasets in a timely manner.

The DeSIRe code is open and publicly available in the GitHub repository\footnote{\url{https://github.com/BasilioRuiz}}.

\section*{Acknowledgements}

C.~Quintero Noda was supported by the EST Project Office, funded by the Canary Islands Government (file SD 17/01) under a direct grant awarded to the IAC on ground of public interest. This activity has also received funding from the European Union’s Horizon 2020 research and innovation programme under grant agreement No 739500. C.~Quintero Noda also acknowledges the support of the ISAS/JAXA International Top Young Fellowship (ITYF) and the JSPS KAKENHI Grant Number 18K13596. This work was supported by ISAS/JAXA Small Mission-of-Opportunity program for novel solar observations and JSPS KAKENHI Grant Number JP18H05234. This work was supported by the Research Council of Norway through its Centres of Excellence scheme, project number 262622, and through grants of computing time from the
Programme for Supercomputing. This work was supported by Fundação para a Ciência e a Tecnologia (FCT) through the research grants UIDB/04434/2020 and UIDP/04434/2020. This work has also been supported by Spanish Ministry of Economy and Competitiveness through the project ESP-2016-77548-C5-1-R and RTI2018-096886-B-C53. D. Orozco Su\'{a}rez also acknowledges financial support through the Ram\'{o}n y Cajal fellowships. CITEUC is funded by National Funds through FCT-Foundation for Science and Technology (project: UID/Multi/00611/2013) and FEDER - European Regional Development Fund through COMPETE 2020 Operational Programme Competitiveness and Internationalization (project: POCI-01-0145-FEDER-006922). NSO is operated by the Association of Universities for Research in Astronomy (AURA), Inc. under cooperative agreement with the National Science Foundation (NSF).

\bibliographystyle{aa} % style aa.bst
\bibliography{inversion} % your references Yourfile.bib

\end{document}